# Pressure-induced reemergence of superconductivity in $BaIr_2Ge_7$ and $Ba_3Ir_4Ge_{16}$ with cage structures

Cuiying Pei[1], Tianping Ying[2*], Yi Zhao[1], Lingling Gao[1], Weizheng Cao[1], Changhua Li[1], Hideo Hosono[3], and Yanpeng Qi[1,4,5*]

1. School of Physical Science and Technology, ShanghaiTech University, Shanghai 201210, China
2. Beijing National Laboratory for Condensed Matter Physics, Institute of Physics, Chinese Academy of Sciences, Beijing 100190, China
3. Materials Research Center for Element Strategy, Tokyo Institute of Technology, Yokohama 226-8503, Japan
4. ShanghaiTech Laboratory for Topological Physics, ShanghaiTech University, Shanghai 201210, China
5. Shanghai Key Laboratory of High-resolution Electron Microscopy, ShanghaiTech University, Shanghai 201210, China

* Correspondence should be addressed to Y. P. Q. (qiyp@shanghaitech.edu.cn) or T. P. Y. (ying@iphy.ac.cn)

**ABSTRACT:**

The clathrate-like or caged compounds have attracted continuing interest due to their structural flexibility as well as fertile physical properties. Here we report a pressure-induced reemergence of superconductivity in $BaIr_2Ge_7$ and $Ba_3Ir_4Ge_{16}$, two new caged superconductors with two-dimensional building blocks of cage structures. After suppressing the ambient-pressure superconducting (SC-I) state, a new superconducting (SC-II) state emerges unexpectedly, with $T_c$ increased to a maximum of 4.4 K and 4.0 K for $BaIr_2Ge_7$ and $Ba_3Ir_4Ge_{16}$, respectively. Combined with high-pressure synchrotron x-ray diffraction and Raman measurements, we propose that the reemergence of superconductivity in the caged superconductors can be ascribed to a pressure-induced phonon softening linked to the cage shrink.

# INTRODUCTION

The recent discovery of LaH$_{10}$ superconductors at high pressure with record-high superconducting transition temperatures $T_c \sim 260$ K has fueled the search for room-temperature superconductivity in compressed superhydrides.[1-3] The three-dimensional clathrate-like structure of H with La atoms filling the clathrate cavities has been described as an extended metallic hydrogen host structure stabilized by the guest electron donor from La. The introduction of electrons into H$_2$ molecules by the guest atom results in a significant contribution of H to the electronic density of states at the Fermi level. As a result, substantial coupling of electrons on the Fermi surface with high-frequency phonons in response to the motion of the H atoms is crucial for promoting superconductivity.[4-7] Actually, clathrate-like superhydrides with alkaline-earth or other rare-earth atoms were also proposed with potential high $T_c$ superconductivity.[8,9] Following the theoretical prediction, experimental progress on the synthesis of these clathrate superhydrides has been remarkable: YH$_6$,[10,11] YH$_9$,[10,12] ThH$_9$,[13] ThH$_{10}$,[13] CeH$_9$,[14,15] CeH$_{10}$,[16] (LaY)H$_6$,[17] (LaY)H$_{10}$,[17,18] and CaH$_6$[19] with high $T_c$ values in the range of 146–253 K are synthesized subsequently. Thus, a new class of high-temperature superconductors with clathrate-like structure was born.

The creation of a distinctive H clathrate structure incorporating metal inside the voids is the key to the unusually high $T_c$ superconductivity. Aside from hydrogen clathrate cages, the interaction between host-guest and interframework is also significant in influencing superhydride superconductivity. Until now, the experimentally obtained high-temperature superconducting clathrate superhydrides only exist above megabar pressure. The ultrahigh pressure required to synthesize and maintain the clathrate superhydrides impedes their experimental investigation. Alternatively, several clathrate or caged compounds show superconducting transition at ambient pressure, which provides a platform to examine the relationship between guest atom and cage unit in a comparatively benign condition.[20-27] In this work, we utilize the *in-situ* high-pressure method to systematically investigate the evolution of superconductivity for two Ba-filled cage compounds $BaIr_2Ge_7$ and $Ba_3Ir_4Ge_{16}$. We discover that the first superconducting phase in both caged compounds is gradually depressed under pressure. At a higher pressure, a pressure-induced superconducting phase dome emerges. There is no structural phase transition revealed by high-pressure synchrotron x-ray diffraction (XRD). The pressure-induced reemergence of superconductivity in $BaIr_2Ge_7$ and $Ba_3Ir_4Ge_{16}$ can be attributed to phonon softening, which is related to the shrink of the cage.

**EXPERIMENTAL SECTION**

Polycrystalline $BaIr_2Ge_7$ and $Ba_3Ir_4Ge_{16}$ were prepared from stoichiometric amounts of high purity elements by argon arc melting and subsequently annealing in evacuated quartz capsules at 1000°C for 20 h.[23] The superconducting transition was confirmed by magnetization measurements using a Magnetic Property Measurement System (MPMS). *In-situ* high-pressure resistivity measurements were conducted on a nonmagnetic diamond anvil cell (DAC) as described elsewhere.[28-32] A piece of nonmagnetic BeCu was used as the gasket. Cubic BN/epoxy mixture layer was inserted between BeCu gasket and electrical leads as an insulator layer. Four Pt foils were arranged according to the van der Pauw method. *In situ* high-pressure x-ray diffraction (XRD) measurements were performed at beamline BL15U of Shanghai Synchrotron Radiation Facility with the x-ray wavelength $\lambda$ = 0.6199 Å. A symmetric DAC with 200 μm culet was used with rhenium gasket. Silicone oil was used as the pressure transmitting medium (PTM) and pressure was determined by the ruby luminescence method.[33] Two-dimensional diffraction images were analyzed using the FIT2D software.[34] Rietveld refinements on crystal structures under high pressure were performed using the General Structure Analysis System (GSAS) and the graphical user interface EXPGUI.[35,36] An *in situ* high-pressure Raman spectroscopy investigation on $BaIr_2Ge_7$ and $Ba_3Ir_4Ge_{16}$ was performed by a Raman spectrometer (Renishaw in-Via, UK) with a laser excitation wavelength of 532 nm and a low-wavenumber filter.

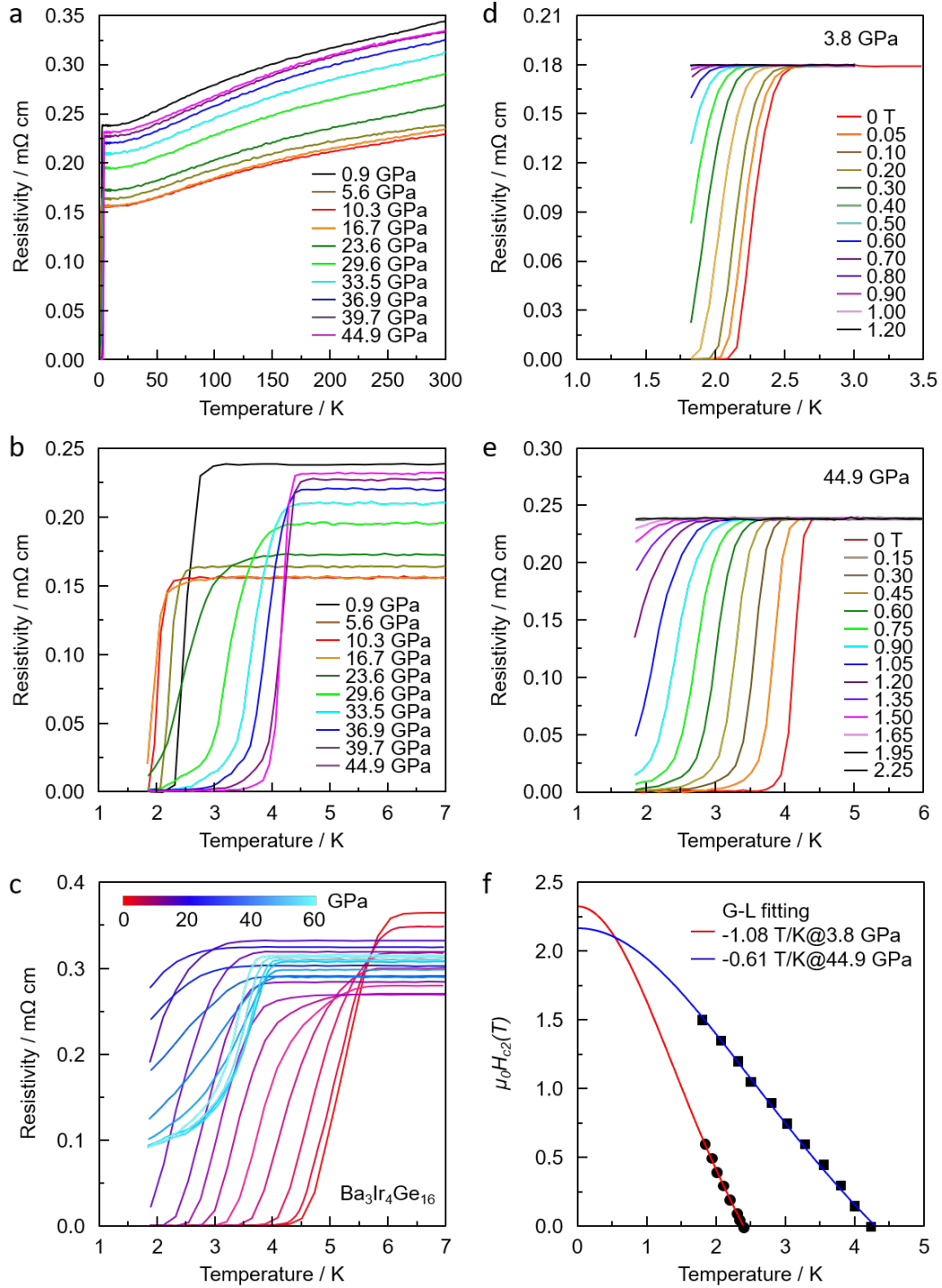

Figure 1. (a) Electrical resistivity of $BaIr_2Ge_7$ as a function of temperature at various pressures. Temperature-dependent resistivity of $BaIr_2Ge_7$ (b) and $Ba_3Ir_4Ge_{16}$ (c) in the vicinity of the superconducting transition. Temperature dependence of resistivity under different magnetic fields for $BaIr_2Ge_7$ at 3.8 GPa (d) and 44.9 GPa (e), respectively. (f) The estimated upper critical field for $BaIr_2Ge_7$. Here, $T_c$ is determined as the 90% drop of the normal state resistivity. The solid lines represent the fits based on the Ginzburg-Landau (G-L) formula.

Table 1. Structure and superconductivity properties of $BaIr_2Ge_7$ and $Ba_3Ir_4Ge_{16}$.

| sample | structure | space group | state | $T_{cmax}$ / K | $H_{c2}$ / T |
|---|---|---|---|---|---|
| $BaIr_2Ge_7$ | orthorhombic | *Ammm* | SC-I | 2.7@0.6 GPa | 2.3@3.8 GPa |
| | | | SC-II | 4.4@39.7 GPa | 2.2@44.9 GPa |
| $Ba_3Ir_4Ge_{16}$ | tetragonal | I4/*mmm* | SC-1 | 5.8@0.1 GPa | 2.0@2.0 GPa |
| | | | SC-II | 4.0@35.2 GPa | 1.7@60.5 GPa |

**RESULTS AND DISCUSSION**

At ambient pressure, both $BaIr_2Ge_7$ and $Ba_3Ir_4Ge_{16}$ exhibit typically metallic behavior and show superconducting transition at 2.7 K and 5.9 K, respectively (Supplemental Information Figure S1).[23-25] Hence, we measured electrical resistivity $\rho(T)$ for both compounds at various pressures. Figure 1a presents the temperature dependence of the resistivity of $BaIr_2Ge_7$ at various pressure up to 44.9 GPa. The resistivity of $BaIr_2Ge_7$ exhibits a non-monotonic evolution with increasing pressure. Over the whole temperature range, the resistivity is first suppressed with applied pressure and reaches a minimum value at about 12 GPa, then displays the opposite trend with further increasing pressure. At lower pressure region, the superconducting transition temperature $T_c$ is suppressed to a minimum of 2.1 K at 16.7 GPa. Surprisingly, the $T_c$ starts to increase rapidly upon further increasing the pressure above 20 GPa, reaching a maximum value of 4.4 K at 39.7 GPa. Further increasing the pressure, the $T_c$ slowly decreases to form a complete dome shape. Compared with 3.2 K at ambient pressure, the $T_c$ at the high-pressure region is much enhanced. The measurements on different samples of $BaIr_2Ge_7$ for independent runs give consistent and reproducible results (Supplemental Information Figure S2), confirming the intrinsic superconductivity under pressure. The pressure-induced reentrant superconductivity resembles the situation in a variety of compounds, including $K_xFe_{2-y}Se_2$,[37] $(Li_{1-x}Fe_x)OHFe_{1-y}Se$,[38] $K_2Mo_3As_3$,[39] $Sr_{0.065}Bi_2Se_3$[40] and $CsV_3Sb_5$[30,41-43]. In addition, we also performed transport measurements for $Ba_3Ir_4Ge_{16}$ under high pressure and a similar evolution of $\rho(T)$ is observed as shown in Figure 1c and Figure S3. In the first superconducting region (SC-I), applying pressure rapidly suppresses the $T_c$ below 2 K at ~20 GPa. In the second superconducting region (SC-II), the $T_c$ continuously increases to the highest 4.0 K at around 30 GPa, which is slightly lower than that in the SC-I region. Beyond this pressure, $T_c$ decreases very slowly and exhibits a typical dome-like feature. The details are summarized in Table 1.

To confirm whether the new resistance drop observed in $BaIr_2Ge_7$ is related to a superconducting transition, we applied a magnetic field to the sample subjected to 3.8 GPa and 44.9 GPa, respectively (Figures 1d and 1e). As shown in Figure 1e, this new resistance drop shifts to a lower temperature with increasing magnetic field and is almost suppressed under the magnetic field of 2.0 T at 44.9 GPa. These results indicate that the sharp resistance drop is a superconducting transition. We also measured the $H$-dependent superconducting transition under $P$ = 3.8 GPa, and plotted the curves in Figure 1d. We extract the field ($H$) dependence of $T_c$ for $BaIr_2Ge_7$ at 3.8 GPa and 44.9 GPa and plot the $H(T_c)$ in Figure 1f. The experimental data are fitted using the Ginzburg-Landau (G-L) formula, which allows us to estimate critical fields $\mu_0 H_{c2}$ ~ 2.3 T and 2.2 T for 3.8 GPa and 44.9 GPa, respectively. Although the $\mu_0 H_{c2}$ obtained here is lower than its corresponding Pauli paramagnetic limit $\mu_0 H_P = 1.84 T_c$, the slope of $dH_{c2}/dT$ is notable different: -1.08 T/K and -0.61 T/K for 3.8 GPa and 44.9 GPa, respectively. Our results suggest that the nature of the pressure-induced reentrant superconducting state may differ from that of the initial superconducting state. A similar evolution of $\mu_0 H_{c2}$ is obtained for $Ba_3Ir_4Ge_{16}$ under various pressures and is shown in Figure S4.

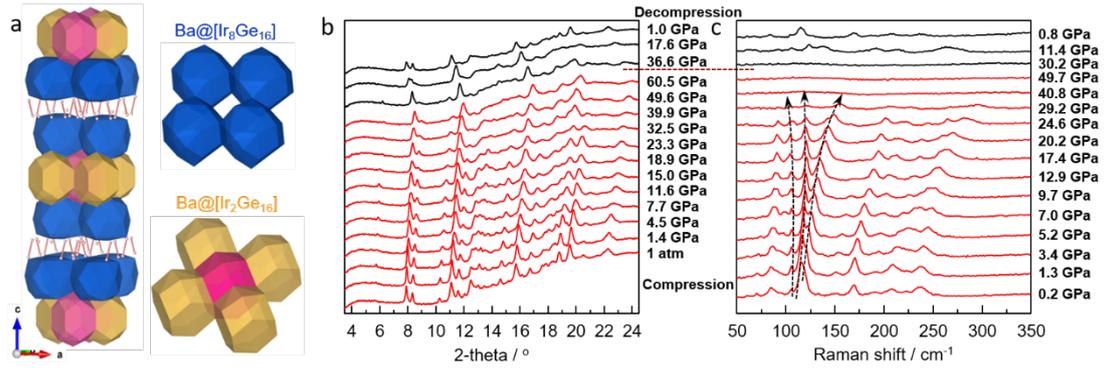

Figure 2. (a) Crystal structures of BaIr$_2$Ge$_7$. Polyhedrons with different colors stand for distinguishing cages. (b) XRD patterns collected at various pressures for BaIr$_2$Ge$_7$ with an x-ray wavelength of $\lambda$ = 0.6199 Å. (c) Selected Raman spectra at various pressure for BaIr$_2$Ge$_7$.

To investigate whether the observed reemergence of superconductivity in pressurized BaIr$_2$Ge$_7$ and Ba$_3$Ir$_4$Ge$_{16}$ is associated with the pressure-induced crystal structure phase transition, we performed *in situ* high-pressure XRD measurements. At ambient pressure, BaIr$_2$Ge$_7$ possesses an orthorhombic structure belonging to the space group, *Ammm*, while Ba$_3$Ir$_4$Ge$_{16}$ crystallizes a tetragonal lattice with space group I4/*mmm* (Figure 2a). Both compounds are composed of 2D networks of cage units, where [Ir$_8$Ge$_{16}$]$^{2-}$ cages are connected by [Ir$_2$Ge$_{16}$]$^{2-}$ cages encapsulating barium atoms. The XRD patterns of BaIr$_2$Ge$_7$ collected at different pressures are shown in Figure 2b. A representative refinement at 0.3 GPa is displayed in Figure S5. All the diffraction peaks can be indexed well to the ambient structure. With increasing pressure, all peaks shift to a higher angle due to the shrinkage of the lattice, and no structure phase transition is observed under pressures up to 60.5 GPa. Structural refinements have been carried out by Rietveld analysis using synchrotron XRD patterns and the unit-cell lattice parameters as a function of pressure have been extracted as depicted in Figure S6c. It is found that the lattice constants shrink obviously at the beginning while decreasing slowly at high pressure. The pressure dependence of volume is shown in Figure 3. With increasing pressure, one can see that the volume decreases but with different slopes below and above a critical pressure $P_c \sim$ 16 GPa. A third-order Birch-Murnaghan equation of state was used to fit the measured pressure-volume (*P-V*) data for BaIr$_2$Ge$_7$.[44] The obtained bulk modulus $K_0$ is 116(2) GPa with $V_0$ = 1545(1) Å$^3$ and $K_0$' = 4. However, the structure becomes less compressible with a higher bulk modulus of 238(3) GPa when the pressure higher than $P_c$. It should be noted that a pressure-induced reentrant superconducting state was observed above critical pressure $P_c$. The *in situ* high-pressure XRD results indicate that the reemergence of superconductivity in $BaIr_2Ge_7$ is not associated with crystal structure phase transition.

We summarize the transport results of $BaIr_2Ge_7$ in a pressure-temperature phase diagram (Figure 3). To confirm the emergence of a second superconducting dome under high pressure, we repeated the measurements with new samples for a second run and proved that all the results are reproducible. The superconducting $T_c$ shows a similar trend with the normal resistivity. The $P$-$T_c$ phase diagram reveals two distinct superconducting regions: the initial superconducting state (SC-I) and the pressure-induced superconducting state (SC-II). In the SC-I region between 1 bar and 16 GPa, $T_c$ is monotonically suppressed with external pressure, and the $T_c$ can be suppressed to 2 K at around 16 GPa. In the SC-II region, $T_c$ increases with pressure and shows a dome shape with the maximum $T_c \sim 4.4$ K at 40 GPa.

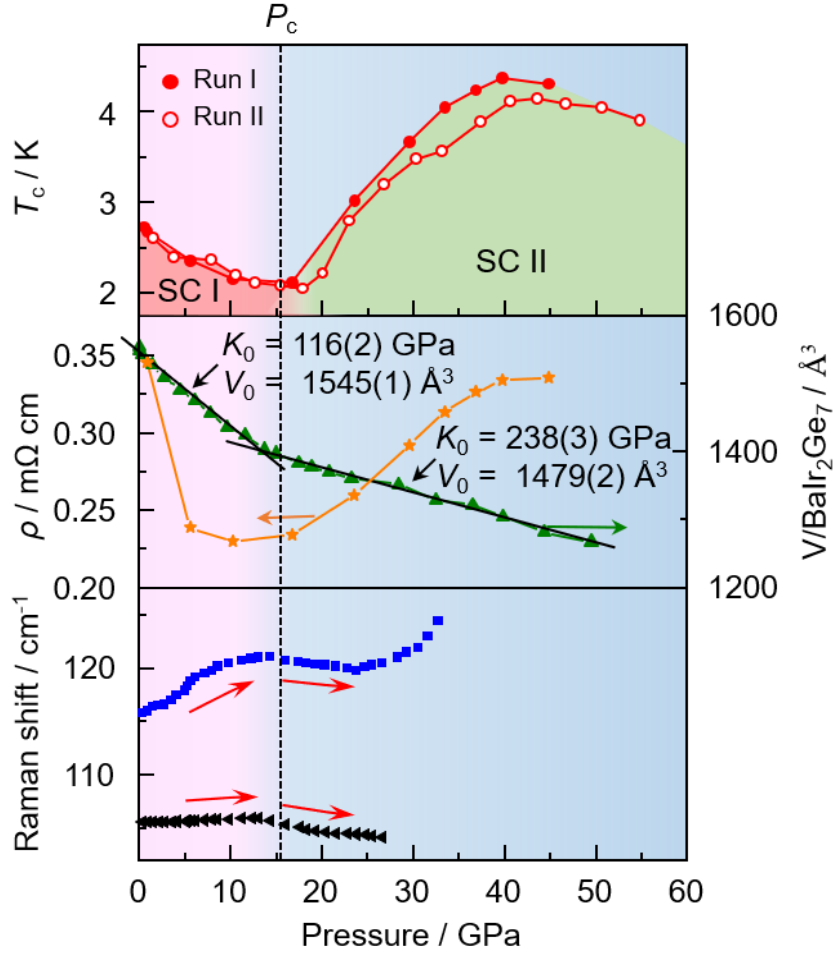

Figure 3. Pressure dependence of the superconducting transition temperatures $T_c$s, resistivity at 300 K, experimental volume relative to *Ammm* phase, and selected Raman to shift for $BaIr_2Ge_7$. The values of $T_c$ were determined from the high-pressure resistivity.

Although pressure-induced reemergence of superconductivity has been reported in various materials, the origin of the second SC dome is still an enigma. Structural phase transition is usually responsible for the two-dome superconductivity. However, since no structural transition was observed from our synchrotron XRD up to 60 GPa, this mechanism can thus be ruled out for $BaIr_2Ge_7$. Another explanation is associated with the competition or coexistence of some order parameters (e.g., CDW, spin density wave, antiferromagnetic order) with superconductivity. Since no charge or magnetic orders are reported in this caged family, charge or antiferromagnetic fluctuations turn out to be irrelevant to the two-dome superconductivity in $BaIr_2Ge_7$. Considering caged structure, the local vibration due to rattling of the Ba atom couples with the low-frequency phonons and conductive electrons may be responsible for superconductivity.

Typically, the low-lying excitation state from the rattling guest atom in the cage can be systemically tuned by external pressure. To gain a more detailed understanding of reemergence superconductivity behavior, we performed high pressure *in situ* Raman spectroscopy measurements on BaIr$_2$Ge$_7$. With increasing pressure, the profile of the spectra remains similar to that at ambient pressure, and the observed modes exhibit a blue shift showing the normal pressure behavior (Figures 2c and S5b). Interestingly, some typical vibration mode (eg. 105.6 cm$^{-1}$ and 115.8 cm$^{-1}$ in ambient condition) display the opposite trend and shows redshift behavior when the pressure is raised up to $P_c$. As summarized in Figure 3, the suppression of superconductivity in SC-I is accompanied by the blue shift of Raman peaks, reaching a minimum $T_c$ and maximum Raman shifts at a turning point of 12 GPa. Further increasing the pressure, both modes of Raman shift steadily decline with the increase of $T_c$ in SC-II. We call for theoretical investigations to determine the origin of these specific vibration modes. Similar behavior is also observed in Ba$_3$Ir$_4$Ge$_{16}$ (Figure S7). Calculations such as electron-phonon couplings (EPC) analysis is very tempting to decipher this anomalous phenomenon. Nevertheless, the present results signify an intimate relationship between the Raman shift and superconductivity, and the pressure-induced phonon softening may be responsible for the reemergence of superconductivity in BaIr$_2$Ge$_7$.

In conclusion, pressure-induced reemergence of the superconductivity is observed in the caged superconductors BaIr$_2$Ge$_7$ and Ba$_3$Ir$_4$Ge$_{16}$. The SC-I is initially suppressed by pressure, and then a second SC dome (SC-II) emerges, with a maximum $T_c$ of ~ 4.4 K and 4.0 K for BaIr$_2$Ge$_7$ and Ba$_3$Ir$_4$Ge$_{16}$, respectively. Synchrotron XRD measurements demonstrate that the reemergence of superconductivity is not associated with any crystal structure phase transition. In combination with *in situ* Raman measurements, our findings suggest that the development of the SC-II state in both caged compounds is a consequence of pressure-induced phonon softening caused by cage shrinkage.

**ACKNOWLEDGMENT**

This work was supported by the National Natural Science Foundation of China (Grant No. U1932217, 11974246 and 12004252), the National Key R&D Program of China

(Grant No. 2018YFA0704300), the Natural Science Foundation of Shanghai (Grant No. 19ZR1477300), the Science and Technology Commission of Shanghai Municipality (Grant No. 19JC1413900) and Shanghai Science and Technology Plan (Grant No. 21DZ2260400). The authors thank the support from Analytical Instrumentation Center (# SPST-AIC10112914), SPST, ShanghaiTech University. The authors thank the staffs from BL15U1 at Shanghai Synchrotron Radiation Facility for assistance during data collection.